\documentclass[aps,pre,amsmath,lengthcheck,superscriptaddress,onecolumn]{revtex4-2}

\usepackage{graphicx}\graphicspath{ {figures/} }
\usepackage{hyperref}
\hypersetup{colorlinks,allcolors=blue,breaklinks}

\newcommand{\be}{\begin{equation}}
\newcommand{\ee}{\end{equation}}
\newcommand{\ba}{\begin{align}}
\newcommand{\ea}{\end{align}}
\newcommand{\bi}{\begin{itemize}}
\newcommand{\ei}{\end{itemize}}

\newcommand{\bla}{bla\\bla\\bla\\bla\\bla}

\begin{document}

\title{Probability distribution of classical observables}

\author{Pierre Naz\'e}
\email{pierre.naze@unesp.br}

\affiliation{\it Universidade Estadual Paulista, 14800-090, Araraquara, S\~ao Paulo, Brazil}

\date{\today}

\begin{abstract}

In this work, I derive the time-dependent probability density function of classical observables using the Hamiltonian mechanics approach, extending the notion of fluctuation theorems for any observables. In particular, the time-dependent probability density function of the thermodynamic work evolving in the switching time is solved as a special case. From such a relation, I prove Jarzynski's equality and Crook's fluctuation theorem. The particular case of a Gaussian distribution for slowly-varying processes is derived as well at the end. 

\end{abstract}

\maketitle

\section{Preliminaries}

Consider a time-dependent driven Hamiltonian $\mathcal{H}_t=\mathcal{H}(q_t,p_t;\lambda(t/\tau))$ with one degree of freedom, such that a driving is performed during a switching time $\tau$ between the points $\lambda_0$ and $\lambda_0+\delta\lambda$.

\section{Time-dependent probability density function of observables}

Consider an observable $\mathcal{O}(q_0,p_0)$ with one degree of freedom. If the system is prepared in the phase space $\Gamma$ with a probability density function $\rho(q_0,p_0)$ and is evolved in time, the observable will have its own probability density function $\rho_{\mathcal{O}_t}(\mathcal{O}_t,t)=\rho_{\mathcal{O}_t}(\mathcal{O}(q_t,p_t),t)$ at time $t$. At the same instant, the probability density associated will be $\rho_t=\rho(q_t,p_t,t)$. Over the evolution in time, the normalization property holds
\be
\int_{\Gamma_{\mathcal{O}_t}}\rho_{\mathcal{O}_t}(\mathcal{O}_t,t)d\mathcal{O}_t=1,
\ee
for all instants $t$. Consider now the probability measure $\mu_t=\mu(q_t,p_t,t)$ associated with $\rho(q_t,p_t,t)$ in the following way
\be
\mu_t(q_t,p_t,t)=\int_{-\infty}^{q_t}\int_{-\infty}^{p_t}\rho(q'_t,p'_t,t)dq'_tdp'_t.
\ee
It holds the probability normalization
\be
\int_{\Gamma_{\mu_t}}d\mu_t=1, 
\ee
also for all instants $t$. To determine $\rho_{\mathcal{O}_t}$, I multiply both normalizations presented before and made a change of variables $(\mathcal{O}_t,\mu_t)\rightarrow (q_t,p_t)$. Therefore
\be
\int_{\Gamma}\rho_{\mathcal{O}_t}(\mathcal{O}(q_t,p_t),t)\left|\frac{\partial \mathcal{O}_t}{\partial q_t}\frac{\partial\mu_t}{\partial p_t}-\frac{\partial \mathcal{O}_t}{\partial p_t}\frac{\partial\mu_t}{\partial q_t}\right|dq_tdp_t=\int_\Gamma\rho(q_t,p_t,t)dq_tdp_t.
\ee
Equalling the integrands of the left and right sides of the equation, one has
\be
\rho_{\mathcal{O}_t}(\mathcal{O}(q_t,p_t),t)=\frac{\rho(q_t,p_t,t)}{\left|\frac{\partial \mathcal{O}_t}{\partial q_t}\frac{\partial\mu_t}{\partial p_t}-\frac{\partial \mathcal{O}_t}{\partial p_t}\frac{\partial\mu_t}{\partial q_t}\right|},
\ee
or also
\be
\rho_{\mathcal{O}_t}(\mathcal{O}(q_t,p_t),t)=\frac{\rho(q_t,p_t,t)}{\left|\{\mathcal{O}_t,\mu_t\}_t\right|}.
\label{eq:key}
\ee
Observe that the observable probability density function per point $\rho_{\mathcal{O}_t}(\mathcal{O}(q_t,p_t),t)$ is positive. Also, since $\rho(q_t,p_t,t)$ satisfy Liouville's theorem, one has
\be
\rho_{\mathcal{O}_t}(\mathcal{O}(q_t,p_t),t)=\frac{\rho_{\mathcal{O}_0}(\mathcal{O}(q_0,p_0))\left|\{\mathcal{O}_0,\mu_0\}_0\right|}{\left|\{\mathcal{O}_t,\mu_t\}_t\right|}=\frac{\rho(q_0,p_0)}{\left|\{\mathcal{O}_t,\mu_t\}_t\right|},
\ee
and 
\be
\frac{d\rho_{\mathcal{O}_t}(\mathcal{O}(q_t,p_t),t)}{dt}=0,\quad \frac{d\left|\{\mathcal{O}_t,\mu_t\}_t\right|}{dt}=0.
\ee
Constructing the set $\mathcal{E}_{\mathcal{O}_t}=\{(q_t,p_t), \mathcal{O}_t=\mathcal{O}(q_t,p_t,t)\}$, the observable probability density function $\rho_{\mathcal{O}_t}$ will be
\be
\rho_{\mathcal{O}_t}(\mathcal{O}_t)=\sum_{(q_t,p_t)\in \mathcal{E}_{\mathcal{O}_t}}\rho_{\mathcal{O}_t}(\mathcal{O}(q_t,p_t,t),t),
\ee
which is nothing more than the multiplication by the probability density function $\rho(q_t,p_t,t)$
\be
\rho_{\mathcal{O}_t}(\mathcal{O}_t)=N\rho(q_t,p_t,t)\rho_{\mathcal{O}_t}(\mathcal{O}(q_t,p_t,t),t),
\ee
where $N$ is a new normalization factor. One can derive a fluctuation theorem for observables now. First, I consider dimensionless observables made by the multiplication of a factor $\beta$. Indeed, using Eq.~\eqref{eq:key}, one has the following fluctuation theorem
\be
\int_{\Gamma_t}\frac{\left|\{\beta\mathcal{O} (q_t,p_t),\mu_t(q_t,p_t)\}_t\right|}{\rho(q_0,p_0)}\rho(q_t,p_t,t)dq_tdp_t=\int_{\Gamma_t}\left|\{\beta\mathcal{O}(q_t,p_t),\mu_t(q_t,p_t)\}_t\right|dq_tdp_t,
\label{eq:ft}
\ee
which can be written as
\be
\int_{\Gamma_{\mathcal{O}_t}}\frac{\left|\{\beta\mathcal{O}_t,\mu_t\}_t\right|}{\rho_0}\rho_{\mathcal{O}_t}(\mathcal{O}_t,t)d\mathcal{O}_t=\int_{\Gamma_t}\left|\{\beta\mathcal{O}(q_t,p_t),\mu_t(q_t,p_t)\}_t\right|dq_tdp_t.
\ee

\section{Time-dependent work probability density function}

Consider that the time-dependent driven system with one degree of freedom is initially weakly coupled to a heat bath at temperature $\beta^{-1}$. When the system starts to evolve, the heat bath is removed from the system. After the switching time $\tau$, the work performed on the system is
\be
W_\tau(q_0,p_0)=\mathcal{H}_\tau(q_\tau(q_0,p_0),p_\tau(q_0,p_0))-\mathcal{H}_0(q_0,p_0),
\ee
Let us use the same reasoning used in the previous section to derive the time-dependent work probability density function. Using a measure $\mu_\tau$ related to the probability density function $\rho(q_\tau,p_\tau,\tau)$, I make the change of variables $(\beta W_\tau,\mu_\tau)\rightarrow (q_0,p_0)$ and compare it with $\rho(q_0,p_0)$. The work probability density function per points is
\be
\rho_{\beta W_\tau}(\beta W_\tau(q_0,p_0),\tau)=\frac{\rho(q_0,p_0)}{\left|\{ \beta W_\tau,\mu_\tau\}_0\right|}.
\ee

\subsection{Jarzynski's equality}

Now let us digress deducting the fluctuation theorem associated. One has
\be
\frac{\left|\{\beta W_\tau(q_0,p_0),\mu_\tau(q_0,p_0)\}_0\right|}{\rho(q_0,p_0)}=\frac{\left|\{\beta W_\infty (q_0,p_0),\mu'_\infty(q_0,p_0)\}_0\right|}{e^{-\beta \mathcal{H}_0}}e^{\beta \Delta F},
\ee
where I used the fact that $\left|\{\beta\mathcal{O} (q_0,p_0),\mu_\tau(q_0,p_0)\}_0\right|$ is invariant over $\tau$ and consider the measure $\mu'_\infty$ associated with $\rho'_\infty=N_{\rho_\infty} \rho_\infty$, where $N_{\rho_\infty}$ the normalization factor associated for the equilibrium state at $\tau=\infty$. Here, $\Delta F$ is the difference of Helmholtz's free energy calculated at the final and initial equilibrium states. Observe also that
\be
\frac{d\left|\{\beta W_\infty,\mu'_\infty\}_0\right|}{d(\beta\mathcal{H}_\infty)}={\rm sign}(\{ \beta W_\infty,\mu'_\infty\}_0)\left\{\beta W_\infty,\frac{d\mu'_\infty}{d(\beta\mathcal{H}_\infty)}\right\}_0=-\left|\{\beta W_\infty,\mu'_\infty\}_0\right|
\ee
since $\rho'(q_\infty,p_\infty,\infty) = e^{-\beta \mathcal{H}_\infty}$ and I suppose that there exists a measure $m$ to make the transformation to variable $(q_0,p_0)\rightarrow (e^{-\beta \mathcal{H}_\infty},m)$. This leads to
\be
\left|\{ \beta W_\tau,\mu'_\tau\}_0\right|= e^{-\beta \mathcal{H}_\tau}.
\ee
Using the above result and the invariance over $\tau$, one can show 
\be
\int_{\Gamma_t}\left|\{\beta W_\tau(q_0,p_0),\mu_t(q_0,p_0)\}_t\right|dq_tdp_t=1.
\ee
Using Eq.~\eqref{eq:ft}, one obtains Jarzynski's equality~\cite{jarzynski1997}
\be
\int_\Gamma e^{-\beta W_\tau^{\rm irr}}\rho(q_t,p_t,t)dq_tdp_t=1,
\ee
where $W_\tau^{\rm irr}=W_\tau-\Delta F$. 

\subsection{Work probability density function}

The final form of the work probability density function is
\be
\rho_{\beta W_{\tau}}(\beta W_{\tau},\tau)=\rho(q_\tau,p_\tau,\tau)e^{\beta W_\tau^{\rm irr}(q_\tau,p_\tau)}.
\label{eq:rhow}
\ee
The trick here is to express $\rho(q_\tau,p_\tau,\tau)$ in terms of the $W_\tau$. I discuss this in the section~\ref{sec:sv}.

\subsection{Crook's fluctuation theorem}

Observe that such an expression~\eqref{eq:rhow} satisfies Crook's fluctuation theorem~\cite{crooks1999}. Indeed
\begin{align}
    \rho_{W_\tau}(W_\tau^+,\tau) & = \frac{\rho(q_0,p_0)}{e^{-\beta W_\tau^{\rm irr +}}}\\
                  & = \frac{e^{-\beta \mathcal{H}_0}}{e^{-\beta F_0}e^{\beta W_\tau^{\rm irr -}}}\\
                  & = \frac{e^{-\beta \mathcal{H}_\tau}e^{\beta W_\tau^{\rm irr +}}}{e^{-\beta F_\tau}e^{-\beta (-W_\tau^{\rm irr -})}}\\
                  & = \rho_{W_\tau}(-W_\tau^-,\tau)e^{\beta W_\tau^{\rm irr +}}.
\end{align}
Also, relating Crook's fluctuation theorem and Eq.~\eqref{eq:rhow}, and using Liouville's theorem, one has
\be
\rho_{W_\tau}(-W_\tau^-,\tau)=\rho(q_\tau,p_\tau,\tau).
\ee
One more time, to have the complete expression it is necessary to express $\rho(q_\tau,p_\tau,\tau)$ in terms of the $W_\tau^-$.

\subsection{Gaussian distribution in slowly-varying processes}
\label{sec:sv}

In this section, I am going to deduce the Gaussian distribution for the work probability distribution for slowly-varying processes, and discuss how one can express $\rho(q_\tau,p_\tau,\tau)$ in terms of $W_\tau$. Observe first that the previous result was derived for a thermally isolated driven system. Since Jarzynski's equation holds for systems in contact with a heat bath, I consider that the derived probability density functions holds here as well. Assuming this we are going to see {\it a posteriori} that it is reasonable assumption.

For quasistatic process the work probability density function is approximately equal to the initial canonical ensemble. In particular, in this regime, all the initial conditions go to the difference of Helmholtz's free energy between the final and initial equilibrium states
\be
\rho_{W_\infty}(W_\infty,\infty)\approx\rho(q_0,p_0)=\delta(W_\infty-\Delta F).
\ee
To generalize such quantity to any switching time, one must preserve the invariance of $\rho(q_0,p_0)$ over $\tau$ in the new expression related to $W_\tau$. Mathematically speaking, this is done by mollifiers $\mathcal{I}(x)$~\cite{friedrichs1944identity}, where
\be
\int_{-\infty}^{\infty} \mathcal{I}(x)dx=1,\quad \lim_{\epsilon\rightarrow 0}\epsilon^{-1}\mathcal{I}(x/\epsilon)=\delta(x).
\ee
Indeed
\be
\int_{\Gamma_0} \rho(q_0,p_0)dq_0dp_0=\int_{\Gamma_t} \rho(q_t,p_t,t)dq_tdp_t\Rightarrow \int_{-\infty}^{\infty} \delta(W_\infty)dW_\infty=\int_{-\infty}^{\infty} \epsilon^{-1}\mathcal{I}( W_\tau/\epsilon)dW_\tau.
\ee
In particular, I choose the following mollifier
\be
\mathcal{I}(W-\Delta F)=\frac{e^{-\frac{\beta^2(W-\Delta F)^2}{2}}}{\sqrt{2\pi/\beta^2}}.
\ee
Also, $\Delta F$ becomes equal to the averaged work $\overline{W}_\tau$. In particular, for slowly-varying processes, where $\tau\gg 1$, the irreversible work is close to zero, so $e^{\beta W_\tau^{\rm irr}}\approx 1$. Also, I choose $\epsilon=\beta^2\sigma_\tau^2$, where $\sigma_\tau^2$ is the variance of the work, which $\sigma_\tau^2\rightarrow 0$ as $\tau\rightarrow \infty$~\cite{yi2021}. One has then
\be
\rho_{W_\tau}(W_\tau,\tau)\approx\frac{1}{\sqrt{2\pi\sigma_\tau^2}}e^{-\frac{(W_\tau-\overline{W}_\tau)^2}{2\sigma^2_\tau}},
\ee
which leads to the expected result of a Gaussian distribution for slowly-varying processes~\cite{seifert2004}. Observe that $\overline{W}_\tau$ and $\sigma^2_\tau$ should be calculated by alternative definitions than by using this distribution.

\section{Conclusions}

In this work, I derived the time-dependent probability density function of classical observables using the Hamiltonian mechanics approach, extending the notion of fluctuation theorems for any observables. In particular, the time-dependent probability density function of the thermodynamic work evolving in the switching time was solved as a special case. From such a relation, I proved Jarzynski's equality and Crook's fluctuation theorem. The particular case of a Gaussian distribution for slowly-varying processes was derived as well at the end.

\bibliography{PDCM.bib}
\bibliographystyle{apsrev4-2}

\end{document}